\documentclass[a4paper,11pt]{article}

\usepackage{jheppub} 

\usepackage[T1]{fontenc} 
\usepackage[cc]{titlepic}

\newcommand{\x}{\psi_{2}(3823)}
\newcommand{\pp}{\pi^0\pi^0}
\newcommand{\PP}{\pi^+\pi^-}
\newcommand{\pip}{\pi^+}
\newcommand{\pim}{\pi^-}

\newcommand{\LL}{\ell^+\ell^-}
\newcommand{\EE}{e^+e^-}

\newcommand{\GG}{\gamma\gamma}

\newcommand{\psip}{\psi(2S)}
\newcommand{\chico}{\chi_{c1}}
\newcommand{\chict}{\chi_{c2}}

\newcommand{\jpsi}{J/\psi}
\newcommand{\piz}{\pi^0}

\newcommand{\ppjpsi}{\pi^+\pi^-J/\psi}

\title{\boldmath Observation of $\EE\to\pp\x$}

\collaboration{The BESIII Collaboration}
\collaborationImg{\includegraphics[height=0.5in]{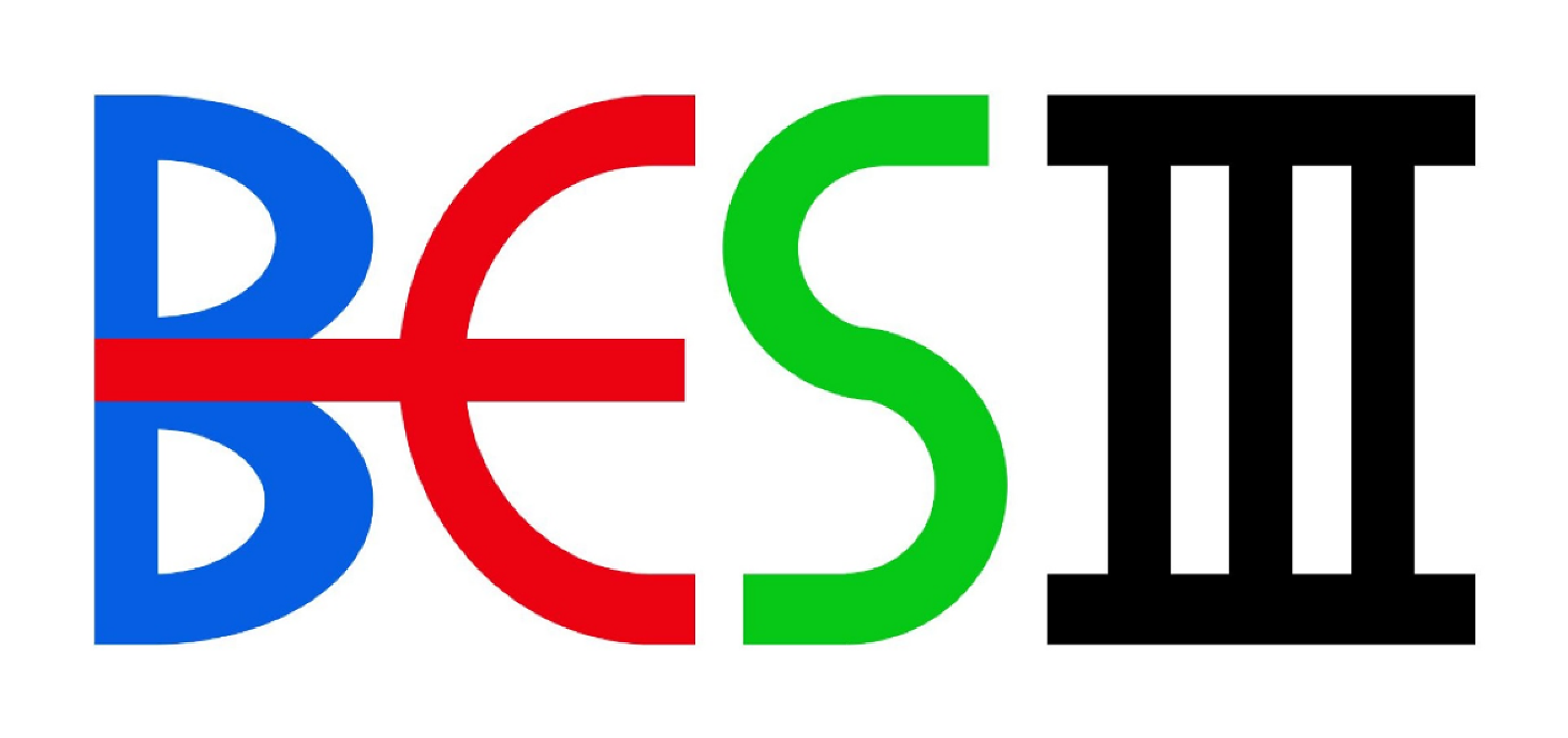}}
\author{
M.~Ablikim$^{1}$, M.~N.~Achasov$^{11,b}$, P.~Adlarson$^{70}$, M.~Albrecht$^{4}$, R.~Aliberti$^{31}$, A.~Amoroso$^{69A,69C}$, M.~R.~An$^{35}$, Q.~An$^{66,53}$, X.~H.~Bai$^{61}$, Y.~Bai$^{52}$
% ...
\\
} 
\affiliation{
$^{1}$ Institute of High Energy Physics, Beijing 100049, People's Republic of China\\
% ...
$^{4}$ Bochum  Ruhr-University, D-44780 Bochum, Germany\\
% ...
$^{11}$ G.I. Budker Institute of Nuclear Physics SB RAS (BINP), Novosibirsk 630090, Russia\\
% ...
$^{31}$ Johannes Gutenberg University of Mainz, Johann-Joachim-Becher-Weg 45, D-55099 Mainz, Germany\\
% ...
$^{35}$ Liaoning Normal University, Dalian 116029, People's Republic of China\\
% ...
$^{52}$ Southeast University, Nanjing 211100, People's Republic of China\\
$^{53}$ State Key Laboratory of Particle Detection and Electronics, Beijing 100049, Hefei 230026, People's Republic of China\\
% ...
$^{61}$ University of Jinan, Jinan 250022, People's Republic of China\\
% ...
$^{66}$ University of Science and Technology of China, Hefei 230026, People's Republic of China\\
% ...
$^{69}$ University of Turin and INFN, (A)University of Turin, I-10125, Turin, Italy; (B)University of Eastern Piedmont, I-15121, Alessandria, Italy; (C)INFN, I-10125, Turin, Italy\\
$^{70}$ Uppsala University, Box 516, SE-75120 Uppsala, Sweden\\
% ...
$^{b}$ Also at the Novosibirsk State University, Novosibirsk, 630090, Russia\\
% ...
}
\emailAdd{besiii-publications@ihep.ac.cn}

\abstract{  
Using a data sample corresponding to an integrated luminosity of 11.3 $\rm fb^{-1}$ collected at center-of-mass energies from $4.23$ to $4.70$~GeV with the BESIII detector, we observe the process $\EE\to \pp\x$ for the first time with a statistical significance of $6.0$ standard deviations.
The ratio of average cross sections for $\EE\to\pp\x$ and $\pip\pim\x$ is determined to be $\mathcal{R}=\frac{\sigma[\EE\to\pp\x]}{\sigma[\EE\to\PP\x]}=0.57\pm 0.14\pm0.05$, which is consistent with expectations from isospin symmetry. 
Here and below, the first uncertainties are statistical and the second are systematic.
The mass of the $\x$ is measured to be $M[\x]=3824.5\pm 2.4\pm 1.0$~MeV/$c^2$.
Due to the limited data sample, an upper limit of $18.8$~MeV at $90\%$ confidence level is set on the intrinsic width of $\x$.
}

\begin{document} 
\maketitle
\flushbottom

\section{Introduction}
\label{sec:intro}
The study of exotic hadrons, whose quark contents are different from conventional baryons and mesons, remains an interesting topic in the field of hadron physics.
In recent years, more than a dozen $XYZ$ particles, which are considered to be good candidates for exotic hadrons~\cite{rmp_olsen,exotics}, have been discovered in the heavy quarkonium energy region.
Here we discuss the $Y$-states, which appear as peaks in the center-of-mass energy dependence of $e^+e^-$ cross sections.
The first candidate, the $Y(4260)$, was discovered by the BaBar experiment in the Initial-State-Radiation (ISR) process $\EE\to\gamma_{\rm ISR}\ppjpsi$~\cite{babar-y4260}, and later confirmed by the Belle experiment in the same process~\cite{belle-cz}. 
By studying the $\EE\to\gamma_{\rm ISR}\pip\pim\psip$ process, the BaBar experiment observed a new resonance, the $Y(4360)$~\cite{babar-psip}.
A detailed study with a larger data sample by the Belle experiment confirmed the $Y(4360)$ resonance, and announced the discovery of a new resonance, the $Y(4660)$~\cite{belle-xl}.
An updated measurement by the BaBar experiment later confirmed the $Y(4660)$ resonance~\cite{babar-y4660}. 
Since the $Y$-states are produced in direct $\EE$ annihilation or via its ISR process, they have the quantum numbers $J^{PC}=1^{--}$, i.e., they are vector states.
From the potential model, the vector charmonium states above the open-charm threshold are expected to decay dominantly to $D^{(*)}\bar{D}^{(*)}$ pairs~\cite{potential}. 
However, as discussed above, these vector $Y$-states are widely discovered in hidden-charm final states, which indicates that they might be exotic states. 
To better understand their underlying nature, more experimental observations are desirable.
Recently, the BESIII experiment reported the observation of resonance structures in the cross section measurement of the process $\EE\to\PP\x$~\cite{ppx3823}, which
suggests that the $\x$ can be used as a probe for the study of $Y$-states.

Since its discovery in 1974~\cite{discovery_jpsi}, the charmonium system has been considered an ideal environment in which to test quantum chromodynamics (QCD) in the non-perturbative regime~\cite{Brambilla:2010cs}.
Just above the open-charm threshold, the $J=2$ member of the $D$-wave spin-triplet, the $1^3D_2$ charmonium state also known as the $\x$, was studied by the E705, Belle, BESIII and LHCb experiments~\cite{E705,belle-3d2,full-rec,LHCb-X3823}.
A recent measurement of the $\x$ mass was reported by the BESIII experiment in the process $\EE\to\PP\x$ ~\cite{ppx3823}.
More detailed measurements of $\x$ properties were also performed at BESIII~\cite{bes3-zhangjl}, but only an evidence of $4.3$ standard deviations was found for the isospin neutral production process $\EE\to\pp\x$.
Moreover, there is still no direct measurement for the $J^P$ of the $\x$.
Further experimental constraints on its quantum numbers would improve our understanding of the $\x$ and charmonium spectroscopy above open-charm threshold. 

In this article, we perform a search for the process $\EE\to\pp\x$ at BESIII by employing a partial-reconstruction method with a signal efficiency which is much higher than that of Ref.~\cite{bes3-zhangjl}. 
The ratio of average cross sections for $\EE\to\pp\x$ and $\EE\to\PP\x$, and the resonance parameters of the $\x$ are measured.
The data sample, corresponding to an integrated luminosity of 11.3 $\rm fb^{-1}$, was taken at center-of-mass (CM) energies from $\sqrt{s}=4.23$ to $4.70$~GeV~\cite{lum} with the BESIII detector~\cite{bes3-detector} operating at the Beijing Electron Positron Collider (BEPCII)~\cite{Yu:IPAC2016-TUYA01}. 
The $\x$ candidate is reconstructed via its decay to $\gamma\chico$, with $\chico\to \gamma J/\psi$ and $J/\psi\to \LL$ ($\ell=e$ or $\mu$). 
The $\pi^0$ candidate is reconstructed via its decay to $\gamma\gamma$.

\section{BESIII detector and Monte Carlo simulation}


The BESIII detector is a magnetic spectrometer located at the BEPCII collider~\cite{Yu:IPAC2016-TUYA01}. 
The cylindrical core of the BESIII detector covers $93\%$ of the full solid angle and consists of a helium-based multilayer drift chamber (MDC), a plastic scintillator time-of-flight system (TOF), and a CsI (Tl) electromagnetic calorimeter (EMC), which are all enclosed in a superconducting solenoidal magnet providing a 1.0~T magnetic field. 
The solenoid is supported by an octagonal flux-return yoke with resistive plate counter muon identification modules interleaved with steel. 
The charged-particle momentum resolution at $1~{\rm GeV}/c$ is $0.5\%$, and the $dE/dx$ resolution is $6\%$ for the electrons from Bhabha scattering. 
The EMC measures photon energies with a resolution of $2.5\%$ ($5\%$) at $1$~GeV in the barrel (end cap) region. 
The time resolution of the TOF barrel part is 68~ps, while that of the end cap part is 110~ps. 
The end cap TOF system was upgraded in 2015 with multi-gap resistive plate chamber technology, providing a time resolution of 60~ps~\cite{etof}.


A {\sc geant4}-based~\cite{geant4} Monte Carlo (MC) simulation software package is used to optimize event selection criteria, determine detection efficiency, and estimate background.
We generate 100000 $\EE\to \pp\x$ signal MC events at each CM energy using an {\sc evtgen}~\cite{evtgen} phase-space model.
The ISR is simulated with {\sc kkmc}~\cite{kkmc}, where the cross section of the $\EE\to\PP\x$ process~\cite{ppx3823} is used as the line shape input.
The maximum ISR photon energy is set corresponding to the 4.1~GeV/$c^2$ production threshold of the $\pp\x$ system.
Final-State-Radiation is handled with {\sc photos}~\cite{photos}.


The background contributions are investigated using an inclusive MC sample, which includes the production of open-charm processes, the ISR production of vector charmonium(-like) states, and the continuum processes incorporated in {\sc kkmc}.
All particle decays are modelled with the {\sc evtgen}~\cite{evtgen} using branching fractions taken from the Particle Data Group~\cite{pdg} when available, or otherwise modelled with {\sc lundcharm}~\cite{lundcharm}.

\section{Event selection and background study}

We select events with two oppositely charged tracks in the polar angle region $|\cos\theta|<0.93$, where $\theta$ is defined with respect to the $z$-axis (the symmetry axis of the MDC). 
For each charged track, the distance of closest approach to the interaction point must be less than $10$~cm along the beam direction and $1$~cm in the plane perpendicular to the beam direction.
Charged tracks with momentum greater than $1.0$~GeV/$c$ are assigned lepton hypotheses.
We make use of the energy depositions in the EMC to identify muons and electrons. 
The deposited energy in the EMC is required to be less than $0.4$~GeV for a muon candidate, while it has to be greater than $1.1$~GeV for an electron. 

Electromagnetic showers identified as photon candidates must satisfy fiducial shower quality and timing requirements ($0\le t \le 700$~ns). 
The minimum energy in the EMC is 25~MeV for barrel showers ($|\cos\theta|<0.80$) and 50~MeV for end cap showers ($0.86<|\cos\theta|<0.92$). 
To exclude showers that originate from
charged tracks,
the angle subtended by the EMC shower and the position of the closest charged track at the EMC
must be greater than 10 degrees as measured from the interaction point. 
We introduce a partial-reconstruction strategy which has a significantly improved efficiency compared to that of the Ref.~\cite{bes3-zhangjl}.
In this strategy, we require at least five photons to be reconstructed in each event ($N_{\gamma}\ge 5$), allowing one missing photon ($\gamma_{\rm miss}$).
The $\gamma_{\rm miss}$ can be any one of the six signal photons. 
The momentum of $\gamma_{\rm miss}$ is determined from momentum conservation.
In addition, we require the number of photons to be $N_{\gamma}\le 6$ to suppress the background contribution from the process $\pp\psip\to\pp\pp\jpsi$.

We apply a four-constraint (4C) kinematic fit to the selected events. 
The invariant mass of the pair of leptons is constrained to the mass of the ${\jpsi}$ ($m_{\jpsi}$),
the mass of the missing photon is constrained to zero,
the invariant mass of a pair of photons is constrained to the mass of the ${\piz}$ ($m_{\piz}$), and the same for another pair of photons.
The values for $m_{\jpsi}$ and $m_{\piz}$ are taken from the PDG~\cite{pdg}. 
Since there is more than one possible combination within an event when selecting the photons and reconstructing the $\piz$s, we retain the one with the minimum $\chi^2$ from the kinematic fit. Events with $\chi^2<15$ are selected for further analysis. 

The two remaining photons other than those used for the reconstruction of the $\piz$s are boosted to the CM frame of the $\x$, and the one with a lower energy is considered to originate from the $\x$ decay, while the other one (with a higher energy) together with the $\jpsi$ candidate is used to reconstruct the $\chico$. The mass window of the $\chico$ candidate is defined as $3.49<M(\gamma\jpsi)<3.53$~GeV/$c^2$.

A study of the inclusive MC sample~\cite{topo} shows that background contributions come from the processes $\EE\to\eta\jpsi$ with $\eta\to \pp\piz$ and $\EE\to\pp\psip$ with $\psip\to\pp\jpsi$.
Background contribution from $\EE\to\eta\jpsi\to\piz\piz\piz\jpsi$ process is effectively rejected by the invariant mass requirement $M(\GG\pp)>0.70$~GeV/$c^2$.
Background contribution from $\EE\to\piz\piz\psip\to\pp\pp\jpsi$ process can be suppressed by vetoing the events in the invariant mass region $3.665<M(\pp\jpsi)<3.700$~GeV/$c^2$.
The contributions from other sources, such as $\EE\to\pp\piz\pip\pim$, are found to be relatively small ($\sim 10\%$ of the total contribution).
The total simulated background only produces a flat distribution in the $\x$ signal region, as shown by the green filled histogram in Figure ~\ref{X-fit}.

\section{Measurements of the mass and width of the $\x$}

\begin{figure}
\begin{center}
\includegraphics[height=2.2in]{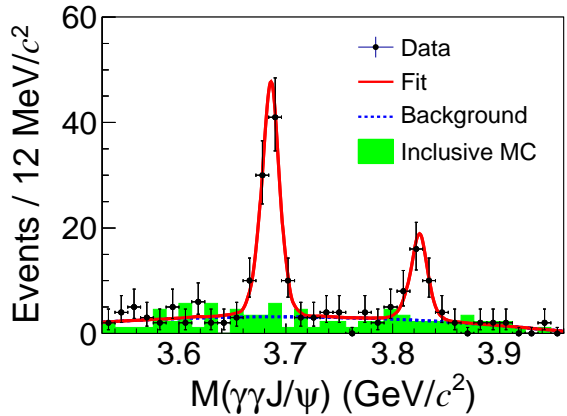}
\caption{The fit to the $M(\GG\jpsi)$ distribution for the data from $\sqrt{s}=4.23$ to $4.70$~GeV.
The dots with error bars are the data. The red solid curve is the total fit. The blue dashed curve is the background in the fit and the green filled histogram is the normalized background from the inclusive MC sample.
}
\label{X-fit}
\end{center}
\end{figure}

Figure~\ref{X-fit} shows the $M(\GG\jpsi)$ distribution for the data from $\sqrt{s}=4.23$ to $4.70$~GeV after the above selection criteria, where prominent $\psip$ and $\x$ signal peaks are observed.
An unbinned maximum likelihood fit is performed to extract the parameters of the $\x$ state.
The probability density function (PDF) of the signal is represented by the sum of the $\psip$ and $\x$ shapes obtained from the MC simulation, each of which is convolved with a Gaussian function to account for the small differences in mass resolution between data and MC simulation.
The parameters of the $\psip$ resonance in simulation are taken from the PDG~\cite{pdg}. 
The mass of the $\x$ in simulation is set to $3823.0$~MeV/$c^2$ and its width is set to zero.
The fit parameter $\sigma$ corresponding to the resolution in Gaussian functions is common for both resonances, while the parameters $\mu_{\psip}$ and $\mu_{\x}$ describing the mass shifts are free.
The background shape is parameterized as a second-order polynomial function.
At BESIII, the $\jpsi$ mass reconstructed by $\LL$ can deviate from PDG value by a level of 0.5~MeV to 3~MeV~\cite{lum}, which is mainly due to calibration, resolution and Final-State-Radiation etc.
To avoid its impact to our mass measurement, the $M(\GG\jpsi)$ mentioned above is defined as $M(\GG\jpsi)\equiv M(\GG\LL)-M(\LL)+m(\jpsi)$, where $m(\jpsi)=3.097$~GeV/$c^2$ is taken from PDG~\cite{pdg}.
The deviation of $\jpsi$ mass therefore partly cancels and the $\x/\psip$ masses are better measured.
In order to further cancel the calibration effects from the two photons, we measure the $\x$ mass with respect to the $\psip$ mass.
Assuming $M[\x]$ and $M[\psip]$ are the true masses of $\x$ and $\psip$, respectively, we calculate their mass difference as
\begin{equation}
M[\x]-M[\psip]=[M[\x]_{\rm input}+\mu_{\x}]-[M[\psip]_{\rm input}+\mu_{\psip}],
\label{mass-diff}
\end{equation}
where $M[\psip]_{\rm input}=M[\psip]=3686.097$~MeV/$c^2$~\cite{pdg}.
The equation then can be derived as
\begin{equation}
M[\x]=M[\x]_{\rm input}+\mu_{\x}-\mu_{\psip}.
\label{mass}
\end{equation}
According to the fit, $\mu_{\x}=(1.8\pm2.4)$~MeV/$c^2$ and $\mu_{\psip}=(0.3\pm1.2)$~MeV/$c^2$. Therefore, by using Eq.~\ref{mass}, the $\x$ mass is measured to be $M[\x]=(3824.5\pm2.4\pm1.2)$~MeV/$c^2$.
Here, the first uncertainty in $M[\x]$ is statistical, which is the uncertainty of $\mu_{\x}$.
The second uncertainty in $M[\x]$ is the uncertainty of $\mu_{\psip}$, which is considered to be systematic, since we take the $\psip$ mass as a reference when measuring the $\x$ mass.
We additionally employ $\psip\to\gamma\chict$ and $\psip\to\eta\jpsi$ data events to increase the $\psip$ data sample. This reduces the uncertainty of $\mu_{\psip}$ in the fit, which gives $\mu_{\psip}=(0.3\pm0.9)$~MeV/$c^2$. 
A more accurate $\x$ mass $M[\x]=(3824.5\pm2.4\pm0.9)$~MeV/$c^2$ is therefore achieved.

The total number of $\x$ candidates extracted from the fit is $N_{\rm total}=30.3\pm6.8$.
A $\chi^2$-test to the fit quality gives $\chi^2/\rm{ndf}=12.13/26=0.47$.
According to Wilks's theorem~\cite{wilkTheorem}, the statistical significance of the $\x$ signal is estimated to be $6.0$ standard deviations, by comparing the difference between the log-likelihood values [$\Delta(-2\ln\mathcal{L})=40.6$] with and without $\x$ signal in the fit, and taking into account the change of the number of degrees of freedom ($\Delta {\rm ndf}=2$).

In order to estimate the width of the $\x$, we slightly modify the fit function described above.
We replace the PDF of the $\x$ signal with a floating-width Breit-Wigner function convolved with Gaussian functions to account for resolution effects.
The parameters of the Gaussian functions are fixed according to the study of the resolution in MC simulation, and the resolution difference between data and MC simulation. 
The $\x$ width is measured to be $\Gamma[\x]=(2.9\pm5.9)$~MeV, corresponding to an upper limit of $18.8$~MeV at the $90\%$ confidence level (including the systematic uncertainty from background shape).
Here the upper limit is set based on the Bayesian method~\cite{pdg}.
The measured mass and width of the $\x$ are consistent with the previous measurements by the BESIII~\cite{full-rec,ppx3823} and LHCb~\cite{LHCb-X3823} experiments.

\section{Measurement of the ratio of average cross sections}

Due to the limited data sample, cross sections at each CM energy cannot be effectively measured.
Instead, the average cross sections for $\EE\to\pp\x$, denoted by $\sigma_{\rm ave}$, and $\EE\to\PP\x$, denoted by $\sigma^{'}_{\rm ave}$, are measured as:
\begin{equation}
\sigma^{(')}_{\rm ave} = \frac{\sum_{i}\sigma^{(')}_i\mathcal{L}_{i}(1+\delta_{i})\epsilon^{(')}_{i}}{\sum_{i}\mathcal{L}_{i}(1+\delta_{i})\epsilon^{(')}_{i}} = \frac{N^{(')}_{\rm total}}{\sum_{i}\mathcal{L}_{i}(1+\delta_{i})\epsilon^{(')}_{i}}\frac{1}{\mathcal{B}^{(')}},
\label{xs-def}
\end{equation}
where $N_{\rm total}$ is the total number of observed $\x$ candidates;
$\sigma_{i}$, $\mathcal{L}_{i}$, $(1+\delta_{i})$, and $\epsilon_{i}$ are the cross section, luminosity~\cite{lum}, radiative correction factor, and efficiency at the $i$-th CM energy point, respectively~(cf. Table~\ref{X-sec});
and $\mathcal{B}$ is the product of branching fractions for the chain of decays involved in each process.  The unprimed variables are for 
$\EE\to\pp\x$ and the primed variables are for $\EE\to\PP\x$ with the values taken from Ref.~\cite{ppx3823}.  Note that the luminosities and radiative correction factors are the same for both processes (the minor difference between the numbers in Table~\ref{X-sec} and that from Ref.~\cite{ppx3823} is due to fluctuation from the size of MC samples).
The ratio of average cross sections, $\mathcal{R}\equiv \sigma_{\rm ave} / \sigma^{'}_{\rm ave}$, is then calculated as
\begin{equation}
\mathcal{R}=\frac{N_{\rm total}}{N_{\rm total}^{'}}\frac{\sum_{i}\mathcal{L}_{i}(1+\delta_{i})\epsilon_{i}^{'}}{\sum_{i}\mathcal{L}_{i}(1+\delta_{i})\epsilon_{i}}\frac{1}{\mathcal{B}^{2}(\piz\to\GG)},
\label{xs-ratio}
\end{equation}
where $\mathcal{B}(\piz\to\GG)$ is the branching fraction of $\piz\to\GG$, and all other branching fractions cancel in the ratio.
According to Eq.~\ref{xs-ratio}, we determine the ratio of average cross sections to be $\mathcal{R}=0.57\pm0.14$ (the uncertainty is statistical), which is consistent with expectations from isospin symmetry ($\mathcal{R}=0.5$), within uncertainty.

\begin{table}
\begin{center}
\caption{The luminosity~\cite{lum}, radiative correction factor and efficiency at each CM energy for the $\EE\to\pp\x$ process. The efficiency for the $\EE\to\PP\x$ process ($\epsilon'$) is also quoted~\cite{ppx3823}. The uncertainty on $\epsilon/\epsilon'$ is related to the MC sample size.} 
\label{X-sec}
\begin{tabular}{cccccc}
  \hline\hline
  $\sqrt{s}$~(GeV)~~ & $\mathcal{L}(\rm pb^{-1})$~~ & $(1+\delta)$~~ & $\epsilon$ & $\epsilon'$  \\
  \hline
  4.2263 & 1056.4 & 0.739 & 0.146$\pm0.001$ & 0.309$\pm0.002$ \\
  4.2580 & 828.4 & 0.745 & 0.143$\pm0.001$ & 0.336$\pm0.002$ \\
  4.2866 & 502.4 & 0.746 & 0.141$\pm0.001$ & 0.333$\pm0.002$ \\
  4.3115 & 501.2 & 0.748 & 0.141$\pm0.001$ & 0.345$\pm0.002$  \\
  4.3370 & 505.0 & 0.747 & 0.147$\pm0.001$ & 0.353$\pm0.002$  \\
  4.3583 & 543.9 & 0.749 & 0.150$\pm0.001$ & 0.357$\pm0.002$  \\
  4.3768 & 522.7 & 0.751 & 0.143$\pm0.001$ & 0.336$\pm0.002$  \\
  4.3954 & 507.8 & 0.762 & 0.137$\pm0.001$ & 0.314$\pm0.002$  \\
  4.4156 & 1043.9 & 0.784 & 0.133$\pm0.001$ & 0.313$\pm0.002$  \\
  4.4359 & 569.9 & 0.818 & 0.138$\pm0.001$ & 0.324$\pm0.002$  \\
  4.4671 & 111.1 & 0.882 & 0.141$\pm0.001$ & 0.341$\pm0.002$  \\
  4.5271 & 112.1 & 1.001 & 0.132$\pm0.001$ & 0.331$\pm0.002$  \\
  4.5745 & 48.9 & 1.075 & 0.129$\pm0.001$ & 0.306$\pm0.002$  \\
  4.5995 & 586.9 & 1.066 & 0.128$\pm0.001$ & 0.304$\pm0.002$  \\
  4.6119 & 103.8 & 0.983 & 0.138$\pm0.001$ & 0.318$\pm0.002$  \\
  4.6280 & 521.5 & 0.831 & 0.151$\pm0.001$ & 0.351$\pm0.002$  \\
  4.6409 & 552.4 & 0.741 & 0.166$\pm0.001$ & 0.378$\pm0.002$  \\
  4.6612 & 529.6 & 0.849 & 0.166$\pm0.001$ & 0.379$\pm0.002$  \\
  4.6819 & 1669.3 & 0.985 & 0.152$\pm0.001$ & 0.356$\pm0.002$  \\
  4.6988 & 536.5 & 1.053 & 0.143$\pm0.001$ & 0.333$\pm0.002$  \\
  \hline\hline
\end{tabular}
\end{center}
\end{table}

\section{Systematic uncertainty}

The systematic uncertainties in the $\x$ mass measurement include those from the absolute mass scale, resolution, signal parameterization, and background shape. 
%
In the $\x$ mass measurement, the $\psip$ mass is used to calibrate the absolute mass scale, so the uncertainty of the measured $\psip$ mass is taken as a systematic uncertainty, which is $0.9$~MeV/$c^2$.
%
We change the width of the Gaussian function in the signal PDF by one standard deviation to do the fits, and the largest mass difference to our nominal fit, $0.3$~MeV/$c^2$, is taken as the systematic uncertainty associated to the resolution.
%
The systematic uncertainty from the parameterization of the $\x$ signal is estimated with different width assumptions.
We input a series of $\x$ widths between zero (which is our nominal value) and its upper limit to generate the MC shapes for the signal PDF constructions, then repeat the fits for the mass measurements. 
The largest mass difference to our nominal fit, $0.3$~MeV/$c^2$, is taken as a systematic uncertainty.
%
We vary the background shape from a second-order polynomial with floating parameters to a second-order polynomial whose parameters are fixed according to the fit to the inclusive MC sample. The fitted $\x$ mass difference between these two background assumptions is found to be small ($0.02$~MeV/$c^2$) and can be neglected.
%
Assuming all the sources are independent, we calculate the total systematic uncertainty by adding them in quadrature, 
resulting in $1.0$~MeV/$c^2$ for the $\x$ mass measurement. 
Table~\ref{mass-err} summarizes the systematic uncertainties for the $\x$ mass measurement.
\begin{table}
\begin{center}
\caption{The systematic uncertainties for the $\x$ mass measurement.} \label{mass-err}
\begin{tabular}{cc}
  \hline\hline
  Source & Mass uncertainty (MeV/c$^2$) \\
  \hline
  Absolute mass scale & 0.9 \\
  Resolution & 0.3 \\
  Signal parameterization & 0.3 \\
  Background shape & $<0.1$ \\
  \hline
  Total & 1.0 \\
  \hline\hline
\end{tabular}
\end{center}
\end{table}


The systematic uncertainties in the measurement of the ratio $\mathcal{R}$ include those from the photon efficiency, signal extraction, kinematic fit, MC decay model, $\chico$ mass window and size of MC samples.
%
The systematic uncertainty from the number of good photons requirement $N_{\gamma} \ge 5$ can be estimated by studying the efficiency 
difference between data and MC simulation, which is $3.1\%$.
The systematic uncertainty due to the requirement $N_{\gamma} \le 6$ originates from the fake-photon-rate difference between data and MC simulation, and this difference is estimated to be $0.1\%$ by studying a control sample of $\EE\to\pp\jpsi$.  
%
The background and signal parameterizations as discussed in the $\x$ mass measurement bring $3.6\%$ and $5.0\%$ differences in the $\x$ signal event yields, which are taken as the systematic uncertainties from signal extraction.
%
A track helix parameter correction method is applied to each MC simulated event during the 4C kinematic fit as discussed in Ref.~\cite{kf-correction}. The difference in detection efficiencies with and without the corrections, $1.2\%$, is assigned as the systematic uncertainty associated to the kinematic fit. 
%
The $\x$ state most likely has quantum numbers $J^{PC}=2^{--}$~\cite{ppx3823}, and the $\pp$ system in $\pp\x$ is expected to be dominated by $S$-wave contribution, such as $f_0(500)$. 
According to spin-parity conservation, value of the orbital angular momentum $L$ between $\pp$ and $\x$ is therefore 2.
We perform MC simulation of the $\EE\to\pp\x$ process with $L=2$ between $\pp$ and $\x$.
The efficiency difference between this model and the nominal three-body phase-space model, $1.0\%$, is taken as the systematic uncertainty from the MC decay model.
%
Using a control sample from the process $\EE\to\PP\psip\to\PP\gamma\chico$, we estimate the systematic uncertainty due to the $\chico$ mass window requirement to be $1.1\%$.
%
The uncertainty from the MC sample size is $0.8\%$.
%
We change the input cross section line shape from a two-resonance interpretation to an alternative single-resonance interpretation, as described in Ref.~\cite{ppx3823}, and the variation among the calculated ratios $\mathcal{R}$ is found to be small ($0.3\%$) and can be neglected with respect to the total uncertainty.
%
The systematic uncertainties from luminosity, reconstruction efficiency of the lepton, branching fractions $\mathcal{B}(\chico\to\gamma\jpsi)$ and $\mathcal{B}(\jpsi\to\LL)$ cancel.
%
The uncertainty from the quoted branching fraction $\mathcal{B}^2(\piz\to\GG)$ is small ($0.03\%$~\cite{pdg}) and neglected.
%
The systematic uncertainties inherent only from the charged channel $\EE\to\PP\x$~\cite{ppx3823} are estimated to be $5.3\%$ in total. 
%
Assuming all the sources are independent and there is no correlation between neutral and charged channels for the above systematic uncertainties, we calculate the total systematic uncertainty by adding them in quadrature, resulting in $8.9\%$ for the measurement of the ratio $\mathcal{R}$.
Table~\ref{sec-err} summarizes the systematic uncertainties related for the ratio $\mathcal{R}$ measurement. 

\begin{table}
\begin{center}
\caption{The systematic uncertainties for the measurement of the ratio $\mathcal{R}\equiv \sigma_{\rm ave} / \sigma^{'}_{\rm ave}$ (the values for $\sigma^{'}_{\rm ave}$ are quoted from Ref.~\cite{ppx3823}). The ``-$"$ means this item is not applicable.}
\label{sec-err}
\begin{tabular}{ccc}
  \hline\hline
  Source & Uncertainty for $\sigma_{\rm ave}$ ~~~&~~~ Uncertainty for $\sigma^{'}_{\rm ave}$   \\
  \hline
  Tracking and photon       & 3.1\% (photon)    & 2.0\% (pion)   \\
  Background shape & 3.6\%           & 1.4\%          \\
  Signal parameterization     & 5.0\%           & 3.9\%          \\
  Kinematic fit             & 1.2\%             & 1.7\%          \\
  MC decay model            & 1.0\%             & 1.8\%          \\
  $\chico$ mass window  & 1.1\%             & -      \\
  MC sample size             & 0.8\%             & 0.6\%          \\
  \hline
  Total                     & 7.2\%             & 5.3\%          \\
  \hline\hline
\end{tabular}
\end{center}
\end{table}

\section{Summary and discussion}

In summary, 
by using a data sample corresponding to an integrated luminosity of 11.3~fb$^{-1}$ collected with the BESIII detector at CM energies from $4.23$ to $4.70$~GeV,
the process $\EE\to \pp\x$ is observed for the first time.
The ratio of average cross sections for $\EE\to\pp\x$ over $\EE\to\pip\pim\x$ is measured to be $\mathcal{R}\equiv \sigma_{\rm ave} / \sigma^{'}_{\rm ave}=0.57\pm 0.14\pm0.05$, which agrees with the expectation from isospin symmetry.
Here and below, the first uncertainties are statistical and the second are systematic.
This result supports the di-pion transition of the $Y$-states to $\x$ observed in $\EE\to\pip\pim\x$~\cite{ppx3823}, though currently we do not have enough data to measure the CM energy dependent cross sections of $\EE\to\pp\x$. 

The mass of the $\x$ is measured to be $M[\x]=(3824.5\pm 2.4\pm 1.0)$~MeV/$c^2$, which is in agreement with the previous measurements~\cite{belle-3d2,full-rec,LHCb-X3823,ppx3823}.
Due to the limited data sample, an upper limit is given to the width of $\x$, which is $\Gamma[\x]<18.8$~MeV at the $90\%$ confidence level.
According to an angular distribution study of the $\x$ from the process $\EE\to\PP\x$ at BESIII~\cite{ppx3823}, the $\x$ is likely a state with quantum numbers $J^{PC}=2^{--}$ assuming the $\PP$ system is dominated by $f_0(500)$.
Since the $\rho^0\to\pp$ decay is forbidden, the observation of $\EE\to\pp\x$ thus further confirms that the $\pi\pi$ system in $\EE\to\pi\pi\x$ comes from $f_0(500)$ decay instead of the $\rho^0$.
It therefore supports the $J^{PC}=2^{--}$ assignment for the $\x$. 

\acknowledgments

The BESIII collaboration thanks the staff of BEPCII and the IHEP computing center for their strong support. This work is supported in part by National Key R\&D Program of China under Contracts Nos. 2020YFA0406300, 2020YFA0406400; National Natural Science Foundation of China (NSFC) under Contracts Nos. 11975141, 11635010, 11735014, 11835012, 11935015, 11935016, 11935018, 11961141012, 12022510, 12025502, 12035009, 12035013, 12192260, 12192261, 12192262, 12192263, 12192264, 12192265; the Chinese Academy of Sciences (CAS) Large-Scale Scientific Facility Program; Joint Large-Scale Scientific Facility Funds of the NSFC and CAS under Contract No. U1832207; 100 Talents Program of CAS; Project ZR2022JQ02 supported by Shandong Provincial Natural Science Foundation; The Institute of Nuclear and Particle Physics (INPAC) and Shanghai Key Laboratory for Particle Physics and Cosmology; ERC under Contract No. 758462; European Union's Horizon 2020 research and innovation programme under Marie Sklodowska-Curie grant agreement under Contract No. 894790; German Research Foundation DFG under Contracts Nos. 443159800, Collaborative Research Center CRC 1044, GRK 2149; Istituto Nazionale di Fisica Nucleare, Italy; Ministry of Development of Turkey under Contract No. DPT2006K-120470; National Science and Technology fund; National Science Research and Innovation Fund (NSRF) via the Program Management Unit for Human Resources \& Institutional Development, Research and Innovation under Contract No. B16F640076; STFC (United Kingdom); Suranaree University of Technology (SUT), Thailand Science Research and Innovation (TSRI), and National Science Research and Innovation Fund (NSRF) under Contract No. 160355; The Royal Society, UK under Contracts Nos. DH140054, DH160214; The Swedish Research Council; U. S. Department of Energy under Contract No. DE-FG02-05ER41374.

\end{document}